\documentclass{PoS}
\usepackage{graphicx}
\usepackage{amsmath, amssymb}
\usepackage{dcolumn}
\usepackage{bm}
\usepackage{xspace}
\usepackage{multirow}
\usepackage{booktabs}
\usepackage{longtable}
\usepackage{textcomp}
\usepackage{subfigure}
\usepackage{enumerate}
\input{symblibrary.tex}
\title{ 
       Evidence for the bottom baryon resonance 
       state \( {\Lambda^{*0}_{b}} \) with~the~CDF~II~detector 
      }
\ShortTitle{\( {\Lambda^{*0}_{b}} \) with CDF~II}
\author{\speaker{Igor~V.~Gorelov}\\
        (on behalf of the CDF~Collaboration)\\
        University~of~New~Mexico,~Albuquerque,~USA\\
        E-mail: \email{gorelov@fnal.gov}}
\abstract{
      Using data from \(\proton\antiproton\) collisions at
      \(\sqrt{s}=1.96\tev\) recorded by the CDF~II~detector at the
      Fermilab Tevatron, we present evidence for the excited resonance
      state \(\LbS \) in its fully reconstructed decay mode to
      \(\Lb\pim\pip\) where \(\Lb\to\Lc\pim\) with \(\Lc\to\pKpi\).  The
      analysis is based on a data sample corresponding to an integrated
      luminosity of \({9.6}\invfb\) collected by an online event
      selection based on tracks displaced from the
      \(\proton\antiproton\) interaction point.  The local significance
      of the observed signal is \(4.6\sigma \) while the significance of
      the signal for the search region is \(3.5\sigma\).  The mass of
      the observed state is found to be 
      \(5919.5\,\pm0.35\,\stat\pm1.72\,\syst\,\mevcc\).
}
\FullConference{36th International Conference on High Energy Physics,\\
		July 4-11, 2012\\
		Melbourne, Australia}
\begin{document}
%
%
%
%
%
  Baryons with a heavy quark \( Q \) can be viewed as a useful
  laboratory for quantum chromodynamics (QCD) in its confinement domain.
  An experimental measurement of a new heavy quark baryon state adds
  another constraint in sampling the confinement QCD force with
  experimental data. 
\par 
  The first result on bottom baryon resonances was obtained by CDF with
  the discovery of the \(S\)-wave \Sgbst states in the \(\Lb\pipm\) decay
  modes~\cite{:2007rw}. Recently CDF has confirmed this observation
  presenting the measurements of the masses and widths of the
  \(\Sgbstpm\) baryons~\cite{CDF:2011ac}.
  In this report, we present evidence for the \(P\)-wave bottom
  resonance \LbS, predicted at a mass scale to be next to the
  established \Sgbst baryons. We have searched for candidate \LbS
  baryons with the complete data sample of \(9.6\invfb\).
  Our result provides an additional contribution to
  the currently small number of heavy quark baryon observations.
%
\par 
  The models describing the heavy hadrons in the framework of heavy quark
  effective theories (HQET)~\cite{Neubert:1993mb} treat a heavy
  baryon as a system consisting of a heavy quark \( Q\) considered as a static
  color source with mass \( m_{Q}\gg\LQCD \) and of a light diquark 
  \(qq \) with a gluon field~\cite{Isgur:1989vq}.
  Thence the bottom \b quark and the spinless \( [\u\d] \) diquark make
  the lowest-lying singlet ground state \( J^{P} = {\frac{1}{2}}^{+} \),
  the experimentally well established \Lb baryon~\cite{Beringer:1900zz}.
  When the \( [\u\d] \) diquark acquires an orbital excitation with
  \(L=1 \) relative to the heavy quark \b, the two excited states \LbS
  emerge with the same quark content as a singlet \Lb, with isospin
  \(I=0\) but with total spin \(J^{P} = {\frac{1}{2}}^{-}\) and
  \(J^{P} = {\frac{3}{2}}^{-}\)~\cite{Korner:1994nh}.  These isoscalar
  states are the lowest-lying \(P\)-wave states that can decay to the
  singlet \Lb via strong processes involving emission of a pair of
  soft pions -- given the parity \(P \) is conserved and provided
  sufficient phase space is available.  Both \LbS particles are
  classified as bottom baryon resonant states. 
\par   
  Several recent theoretical predictions on masses of the excited heavy baryons
  \LbS are available~\cite{Karliner:2007cu,Ebert:2005xj,AzizaBaccouche:2001pu}.
  Based on the predictions, the mass difference \(M(\LbS)\,-\,M(\Lb)\)
  for the first, \(J^{P}={\frac{1}{2}}^{-} \) state, is predicted to be
  of \(\sim\,300-310\mevcc\). The mass splitting between different
  \(J^{P}\) states,
  \(M(\LbS,\,J^{P}={\frac{3}{2}}^{-})-M(\LbS,\,J^{P}={\frac{1}{2}}^{-})\),
  is evaluated to be of order \(10-17\mevcc \).
%
%
%
\par 
  The component of the \(\cdf2 \) detector~\cite{Acosta:2004yw} most
  relevant to this analysis is the charged particle tracking
  system. The tracking system operates in a uniform axial magnetic
  field of \(1.4\,{\rm T}\) generated by a superconducting solenoidal magnet.
  The inner tracking system comprises three silicon detectors:
  layer~00~(L00), {the silicon vertex detector}~(SVX~II), and {the intermediate
  silicon layers}~(ISL)~\cite{Sill:2000zz}. 
  A large open cell cylindrical drift chamber, the central outer
  tracker~(COT)~\cite{Affolder:2003ep}, completes the CDF
  detector tracking system. 
  The silicon tracking system provides fine resolution on a transverse
  impact parameter \({d_{0}}\) of
  \({\sigma_{d_{0}}}\simeq{35}\mkm\) (with the \(\approx{28}\mkm \) beam spot
  size included).  The combined track transverse momentum resolution of
  the whole tracking system is
  \({\sigma({\pt})}/{\pt}\simeq{0.07\%}\,{\pt}\,[\gevc]^{-1}\).
\par 
  This analysis relies on a three-level trigger used for the online
  event selection to collect large data samples of multibody hadronic
  decays of \b-flavor states. We refer to this as the displaced
  two-track trigger. The trigger requires two tracks in the COT with
  \(\pt>2.0\gevc \) for each track~\cite{Thomson:2002xp}.
  A further
  requirement that the impact parameter \({d_{0}}\) of each track lie in the range
  \(0.12-1\mm\) makes an effective selection of long-lived
  \b-flavor particles~\cite{Ashmanskas:2003gf}.  Finally, the distance
  \(\lxy\) in the transverse plane between the beam axis and the
  intersection point of the two tracks projected onto their total
  transverse momentum is required to be greater than \(200\mum\). 
%
%
%
\par 
  Using the dataset collected with the displaced two-track trigger,
  we reconstruct the \LbS candidate states in the exclusive strong decay
  \(\LbS\to\Lb\pim_{\mathit{s}}\pip_{\mathit{s}}\) followed by the weak
  decays \(\Lb\to\Lc\pim_{b}\) and \(\Lc\to\pKpi\)~\cite{notation:cc}.
  The analysis of the \LbS mass distributions is performed using the
  \(Q\) value, where
    \(Q = m(\Lb\pim_{\mathit{s}}\pip_{\mathit{s}}) - m(\Lb) - 2\,{m_{\pi}}\,\,,\) 
  \( m(\Lb) \) is the reconstructed \(\Lc\pim_{b}\) mass and 
  \({m_{\pi}} \) is the known charged pion mass.
  The mass resolution of the \Lb signal and most of the systematic
  uncertainties cancel in the mass difference spectrum.
  We search for narrow structures in the \(Q\) value spectrum 
  within the range of \(6 - 45\mevcc \) motivated by the theoretical 
  estimates~\cite{Karliner:2007cu,Ebert:2005xj,AzizaBaccouche:2001pu}.  
%
%
\par  
  The analysis begins with reconstruction of the \(\LcpKpi\) decay
  by fitting three tracks to a common vertex.  
  Standard quality requirements are applied to each track, and only
  tracks with \(\pt>400\mevc \) are used.  All tracks are refit using
  pion, kaon and proton mass hypotheses to properly correct for the
  differences in multiple scattering and ionization energy loss.  No
  particle identification is used in this analysis.
  The invariant mass of the \(\Lc\) candidate is required to be within
  \(\pm18\mevcc \) of the world-average \(\Lc\)
  mass~\cite{Beringer:1900zz}.
  The momentum vector of the \(\Lc\) candidate is then extrapolated to
  intersect with a fourth track that is assumed to be a pion, to
  form the \( \Lb\to\Lc\pim_{b} \) candidate. The \Lb vertex is
  subjected to a three-dimensional kinematic fit with the \(\Lc\)
  candidate mass constrained to its world-average
  value~\cite{Beringer:1900zz}. The probability of the constrained \Lb
  vertex fit must exceed \( 0.01\% \)~\cite{CDF:2011ac}.
 The proton from the \(\Lc\) candidate is required to have
  \(\pt>2.0\gevc \) to contribute to the trigger decision. 
   The momentum criterion for the
  \(\pim_{b} \) from the \( \Lb \) has been optimized by maximizing the
  score function \( S_{\Lb}/(1+\sqrt{B}) \), where
  \(S_{\Lb}\) is the number of \(\Lb\) signal events obtained from the
  fit of the \(\Lc\pim_{b}\) invariant mass experimental spectrum 
  and \({B}\) is the number of events in the sideband region, 
  \(50-90\mevcc\), of the \(\LbS \) \(Q \) value experimental
  spectrum. The sideband region boundaries are motivated by the signal predictions
  in \cite{Karliner:2007cu,Ebert:2005xj,AzizaBaccouche:2001pu}.  The
  sideband spectrum is parametrized by a second order Chebyshev
  polynomial. The requirement of \(\pt(\pim_{b})>1.0\gevc\) corresponds
  to the maximum of the score function. The momentum criteria both for
  proton and \(\pim_{b} \) candidates favor these particles to be the
  two contributing to the displaced two-track trigger decision. 
  To keep the slow pions of \LbS decaying within the kinematic acceptance
  of the CDF track reconstruction, the \Lb candidate must have
  \(\pt(\Lb) \) greater than \(9.0\gevc\). This corresponds to the
  maximum of the score function \( S_{\rm MC}/(1+\sqrt{B}) \), where
  \(S_{\rm MC}\) is the \(\LbS \) signal reconstructed in the MC
  simulation and \({B}\) is the number of events in the previously defined
  sideband region of the \(\LbS \) \(Q\) value spectrum.
\par 
  To suppress prompt backgrounds from the primary interactions, the
  decay vertex of the \Lb is required to be distinct from the primary
  vertex.  To achieve this, cuts on the proper lifetime,
  \(\ct{(\Lb)}>200\mkm \), and its significance,
  \(\ct(\Lb)/\sigma_{\it{ct}}>6.0 \), are applied. The first requirement
  confirms the trigger while the second one is set using MC simulation
  data to be fully efficient for the \LbS signal.
  We define the proper lifetime as
   \( \ct{(\Lb)} = \lxy\,{{m_{\Lb}}\,{c}}/{\pt}\,\,,\) 
  where \({m_{\Lb}} \) is the world-average mass of the  
  \Lb~\cite{Beringer:1900zz}. The primary vertex is determined
  event-by-event when computing this vertex displacement.
  We require the \Lc vertex to be associated with a \Lb decay by
  applying a cut on the proper lifetime \(\ct(\Lc)\), where the corresponding
  quantity \( \lxy(\Lc) \) is calculated with respect to the \Lb
  vertex. The requirement \(\ct(\Lc)>-100\,\mkm \) reduces contributions
  from \Lc baryons directly produced in \(\proton\antiproton\)
  interactions and from the random combination of tracks faking \Lc
  candidates (which may have negative \(\ct(\Lc)\) values).
  To reduce combinatorial background and contributions from partially
  reconstructed decays, we require \Lb candidates to point to the primary
  vertex by requiring the impact parameter \({{d_{0}}(\Lb)}\) not to
  exceed \( 80\,\mkm \). Both latter cuts~\cite{CDF:2011ac} are fully
  efficient for the \LbS signal.
\par 
   
\begin{figure}
\begin{center}
  \includegraphics[width=0.65\textwidth]
  {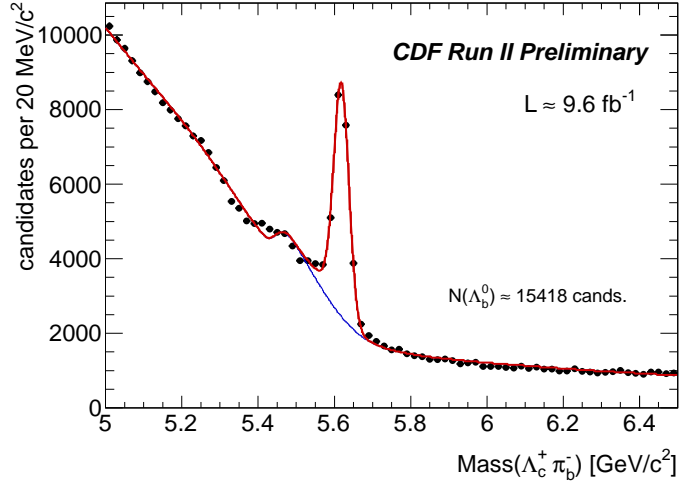}
\caption{ Invariant mass distribution of \(\Lb\to\Lc\pim_{b}\)
          candidates with the projection of a mass fit overlaid.
          The blue line shows the background while the red one 
          corresponds to the signal plus background.
          \label{fig:signal-lb} }
\end{center}
\end{figure}

  Figure~\ref{fig:signal-lb} shows a prominent \Lb signal in the
  \(\Lc\pim_{b}\) invariant mass distribution, reconstructed with the
  criteria explained above. The fit model describing the invariant mass
  distribution comprises the Gaussian \(\Lb\to\Lc\pim_{b}\) signal on
  top of a background shaped by several 
  contributions~\cite{:2007rw,CDF:2011ac,Abulencia:2006df}.  A binned
  maximum-likelihood fit finds a signal of approximately \( 15400 \)
  candidates at the expected \Lb mass, with a signal to background ratio
  around \(1:1\).
%
%
%
\par 
  To reconstruct the \(\LbS\) candidates, each \(\Lb\) candidate with an
  invariant mass within a region of \(5.561-5.677\,\gevcc\) is combined
  with a pair of oppositely charged tracks each assigned to the pion
  hypothesis.  The \(\Lb\) mass region corresponds to an area of
  \(\pm3\sigma\) around the \Lb signal peak as determined by a fit to
  the spectrum of Fig.~\ref{fig:signal-lb}.
  To increase the efficiency for reconstructing \LbS decays near the
  kinematic threshold, the quality criteria applied to soft pion tracks
  are loosened in comparison with those applied to tracks used for the \Lb candidates.
  The basic COT and SVX~II hit requirements are imposed on the 
  \(\pipm_{\mathit{s}}\) tracks, and only tracks with \(\pt>200\mevc \)
  having hits in both trackers and with a valid track fit and error
  matrix are accepted. 
\par  
  To reduce the background level, a kinematic fit is applied to the
  resulting combinations of \(\Lb\pim_{\mathit{s}}\pip_{\mathit{s}} \) candidates 
  to constrain them to
  originate from a common point. The \(\Lb \) candidates are not 
  constrained to a nominal \(\Lb \) mass in this fit.  Furthermore, since the bottom baryon
  resonance originates and decays at the primary vertex, the soft pion
  tracks are required to originate from the primary vertex by requiring
  an impact parameter significance
   \( {{d_{0}}(\pipm_{\mathit{s}})/{\sigma_{d_{0}}}} \) 
  smaller than \(3\)~\cite{:2007rw,CDF:2011ac}. 
  This requirement corresponds to the maximal value of the score
  function \( S_{\rm MC}/(1+\sqrt{B}) \). 
%
%
%
\par
\begin{figure}  
\begin{center}  
  \includegraphics[width=0.66\textwidth]{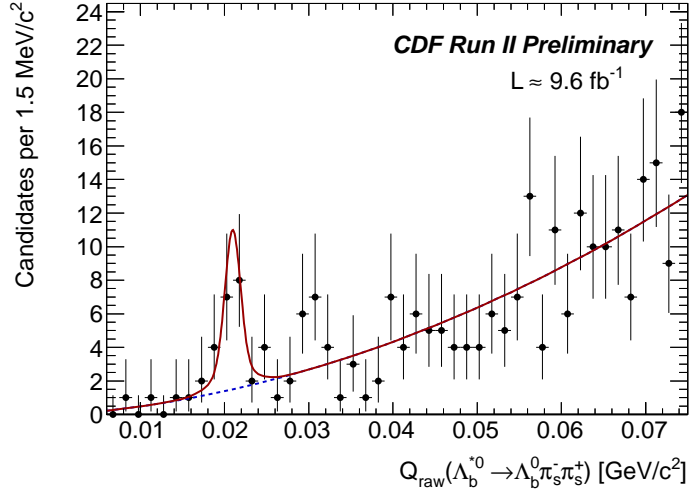}  
  \caption{ The projection of the unbinned fit with a blue line showing
            background only. The binned \(Q\) value distribution of \LbS
            candidates is shown for the range \(0.006\,-\,0.075\gevcc \).
            The soft pion tracks have transverse momentum above \(0.2\gevc\).
          }
\label{fig:signal-lbs}
\end{center}
\end{figure}

  
  The experimental \(\LbS\) \(Q\) value distribution is shown in
  Fig.~\ref{fig:signal-lbs}.  A narrow structure at \(Q\sim21\mevcc\) is
  clearly seen.
  The projection of the corresponding unbinned likelihood fit is
  superimposed on the graph.  The fit function includes a single narrow
  signal structure on top of a smooth background.  The signal is
  parametrized by two Gaussians with the same mean value and with
  their widths and weights set according to Monte Carlo simulation
  studies. The background is described by a second order
  Chebyshev polynomial. The parameters of interest are the position of
  the signal and its yield.
  The negative logarithm of the extended likelihood function is
  minimized over the unbinned set of \(Q\) values observed for the
  candidates in our sample.  The \(Q\) value spectrum is fit over the
  range \(6 - 75\mevcc\).
  The fit finds \(17.3^{+5.3}_{-4.6}\) signal candidates 
  at \(Q = {20.96}\pm{0.35}\,\mevcc\).
\par
  The significance of the signal is determined using a
  \({\log}\)-likelihood ratio statistic~\cite{Wilks,Royall},
  \(D=-2\ln({\mathcal{L}_0}/{\mathcal{L}_1}).\)
  We define hypothesis \({\mathcal{H}_1}\) corresponding to the presence
  of a \( \LbS \) signal on top of the background. 
  The statistic \(D\) is used as a \(\chisq\)
  variable with two degrees of freedom to derive \(p\) values for
  observing a deviation as large as is in our data or larger, assuming
  \({\mathcal{H}_0}\) is true.  Therefore our baseline signal fit has a
  local significance of \(4.6\sigma\). 
  The significance for a \(Q \) search window of \(6 - 45\mevcc\) has
  been determined by running statistical pseudo-experiments in which the
  \({\mathcal{H}_0}\) hypothesis is generated but fit with the
  \({\mathcal{H}_1}\) hypothesis and the corresponding
  \({\log}\)-likelihood ratio statistic is calculated for each trial.
  The fraction of the generated trials having \(D\) above the value
  returned by the fits of the experimental data determines the
  significance. For this case the significance has been found to be
  \(3.5\sigma \).
%
%
%
\par  
  The systematic uncertainties on the mass considered in our analysis
  derive from the CDF tracker momentum scale (the dominant
  contribution); the resolution model described by the sum of two
  Gaussians; and the choice of a background model.
  The uncertainties on the measured mass differences due to the momentum
  scale of the low-\(\pt \) \({\pipm_{\mathit{s}}} \) tracks are
  estimated from the large calibration sample of
  \(\Dstarp\to\Dz\pip_{\mathit{s}} \) events.  The scale factor to be
  applied to the soft pion transverse momentum is found to correct the
  difference between the experimental \(Q \) value in \(\Dstarp \) decays
  and its world-average value~\cite{Beringer:1900zz}. The same factor
  applied for the soft pions in a full Monte-Carlo simulation of
  \(\LbS\to\Lb\pim_{\mathit{s}}\pip_{\mathit{s}} \) decays yields a \(Q\)
  value change of \(-0.28\mevcc \). We take the full value of the
  change as the uncertainty and adjust by \(-0.28\pm0.28\,\mevcc\)
  the \(Q \) value found by the fit of the \(\LbS \) experimental spectrum.
  The systematic uncertainties are summarized in Table~\ref{tab:lbssyst}.
\begin{table}
\begin{center}
\caption{ Summary of systematic uncertainties. }
\label{tab:lbssyst}
\begin{tabular}{lll}
\hline
\hline
{ \bf Source } & {\bf Value}, & {\bf Comment} \\
               & {\mevcc}     & \\
\hline  
  Momentum scale  & \(\pm0.28\) & Propagated from high \\
                  &             & statistics calibration \\
                  &             & \Dstarp sample; \\
                  &             & \(100\% \) of the found  \\
                  &             & adjustment value. \\
  Signal model  & \(\pm0.11\) & MC underestimates \\
                &             & the resolution; choice of \\
                &             & the model's parameters \\
  Background model & \(\pm0.03\) & Consider 3-rd, 4-th power \\
                   &             & polynomials  \\
\hline 
   Total:         & \(\pm0.30\) & Added in quadrature \\
\hline 
\hline
\end{tabular}
\end{center}
\end{table}
%
%
%
\par 
  The analysis results are arranged in Table~\ref{tab:results}.
  From the measured \LbS \(Q\) value we extract the absolute masses
  using the known value of the \(\pi^{\pm}\)
  mass and the CDF  \Lb  mass measurement,
  \(m(\Lb) = 5619.7\pm1.2\,\stat\pm1.2\,\syst\,\mevcc\,\),
  as obtained in an independent sample~\cite{Acosta:2005mq}.
  The \Lb mass statistical and systematic uncertainties contribute to the
  systematic uncertainty  on the \LbS absolute mass.
  The result is closest to the calculation 
  in~\cite{AzizaBaccouche:2001pu}.
  Our result is consistent with the state \(\LbS(5920)\)
  recently observed by the LHCb Collaboration~\cite{Aaij:2012da}.
  The lower production rate of bottom hadrons at the Tevatron
  combined with the low efficiency for soft pion tracks 
  make this sample insensitive to the presence of the \(\LbS(5912)\)
  state observed by LHCb.
\begin{table}
\begin{center}
\caption{ Summary of the final results.  The first uncertainty is 
          statistical and the second is systematic. \label{tab:results}} 
\begin{tabular}{lc}
\hline
\hline
{\bf Value}    & { \( \mevcc \) }   \\
\hline 
\(Q\)   & \(20.68\,\pm\,0.35\stat\,\pm\,0.30\syst \)  \\
\(\Delta M \) & \( 299.82\,\pm\,0.35\stat\,\pm\,0.30\syst \)  \\
\(M(\LbS) \)   & \( 5919.5\,\pm0.35\,\stat\pm1.72\,\syst \)  \\
%
\hline
\hline
\end{tabular}
\end{center}
\end{table}
\par
  In conclusion, we have conducted a search for the
  \(\LbS\to\Lb\pim\pip\) resonance state in its \(Q\) value spectrum,
  and a narrow structure has been identified. The narrow structure has a
  local significance of \(4.6\sigma\) and is interpreted as evidence for
  a \(\LbS \) signal.  The significance of the signal for the search
  region of \(6 - 45\,\mevcc\) is \(3.5\sigma \).
  Our result confirms the state \(\LbS(5920)\) 
  observed by the LHCb Collaboration~\cite{Aaij:2012da}.
\begin{acknowledgments}
  The author is grateful to his colleagues from the CDF {\it B}-Physics
  Working Group for useful suggestions and comments made during
  preparation of this talk.
\end{acknowledgments}

\begin{thebibliography}{99}
\bibitem{notation:cc}
  Low momentum pions produced near the kinematic threshold in strong
  decays of the \(\LbS \) are indicated as \(\pipm_{\mathit{s}}\).
  \(\pim_{b}\) designates a pion produced in the weak decay of the
  \(\Lb\).  Unless otherwise stated all references to a specific charge
  combination imply the charge conjugate combination as well.
  Specifically, \(\LbSBar\to\LbBar\pip_{\mathit{s}}\pim_{\mathit{s}}\),
  \( \LbBar\to\Lcbar\pip_{b} \), \(\Lcbar\to\overline{\proton}\Kp\pim
  \), and \(\Dstarm\to\Dzb(\to\Kp\pim)\pim_{\mathit{s}}\).
\end{thebibliography}
\end{document}